\documentclass[prl,twocolumn,aps,showpacs]{revtex4}
\usepackage{graphicx}
\usepackage{dcolumn}
\usepackage{bm,bbm}
\usepackage{color}
\usepackage{amsmath}
\usepackage{amsfonts}
\usepackage{amssymb}
\usepackage{relsize}

\usepackage{ulem}

\newcommand{\beq}{\begin{equation}}
\newcommand{\eeq}{\end{equation}}
\newcommand{\bqa}{\begin{eqnarray}}
\newcommand{\eqa}{\end{eqnarray}}

\newcommand{\ro}[1]{\left( {#1} \right)}
\newcommand{\an}[1]{\left\langle{#1}\right\rangle}

\newcommand{\red}{\color{red}}

\newcommand{\blk}{\color{black}}
\newcommand{\blu}{\color{blue}}
\definecolor{ngreen}{rgb}{0.2,0.6,0.2}
\newcommand{\grn }{\color{ngreen}}
\definecolor{golden}{rgb}{0.8,0.6,0.1}
\newcommand{\gold}{\color{golden}}

\renewcommand\sout[1]{}

\renewcommand\blu{\blk}
\renewcommand\red{\blk}
\renewcommand\gold{\blk}
\renewcommand\grn{\blk}

\begin{document}
\title{Experimental Test of Universal Complementarity Relations}
\author{Morgan M. Weston, Michael J. W. Hall, Matthew S. Palsson, Howard M. Wiseman and Geoff J. Pryde}
\affiliation{Centre for Quantum Computation and Communication Technology (Australian Research Council), Centre for Quantum Dynamics, Griffith University, Brisbane, QLD 4111, Australia}

\begin{abstract}
Complementarity restricts the accuracy with which incompatible quantum observables can be jointly measured.  
Despite popular conception, the  Heisenberg uncertainty relation does not quantify this principle. 
  We report the  experimental verification \blk of universally valid  complementarity relations,  
    including an  improved 
    relation derived here. 
We exploit Einstein-Poldolsky-Rosen correlations between two  photonic  qubits, to jointly measure incompatible observables of one. 
The   product of our  measurement inaccuracies   is low enough \blk to violate   the \blk widely used, 
but not universally valid,  Arthurs-Kelly relation. 
\end{abstract}


\maketitle

{\it Introduction.}---Niels Bohr repeatedly emphasised that the fundamental distinction between quantum and classical mechanics is the principle of complementarity,  which states  that the experimental arrangements required to accurately measure two different observables are, in general, incompatible \cite{bohr1}. 
 Indeed, in replying to the famous critique by Einstein, Podolsky and { Rosen on} the completeness of the quantum theory \cite{epr},  Bohr stated that {\it ``it is only the mutual exclusion of any two experimental procedures, permitting the unambiguous definition of complementary physical quantities, which provides room for new physical laws''} \cite{bohrepr}.

It is commonly thought that Bohr's complementarity principle is captured by Heisenberg-type uncertainty relations such as   $\Delta Q  \Delta P \geq \hbar/2$, where   $\Delta Q$   \grn ($\Delta P$) is \blk the predicted \grn spread  of  
 a perfect position (momentum) measurement, \blk arising from the underlying position (momentum) spread of a particle's wave function.
 However, this relation does not reflect \grn the impossibility of 
 performing \blk such perfect measurements at the same time.  \blu Nor do measurement-disturbance relations \cite{ozawadist,ozawaannphys, ozawanat,rozema}, since these refer to 
sequential rather than joint measurements. \blk
By  contrast, a    \textit{complementarity relation} 
 should  restrict the degree of accuracy with which 
position and momentum  --- or indeed any two  incompatible quantities --- 
can be measured simultaneously. 

 In particular, whereas two incompatible \blk observables $A$ and $B$ are not jointly measurable  with perfect accuracy, \blk  one may   still 
\blu make {\it inaccurate} joint estimates, \blk 
 $A_\textrm{est} \approx A$ and $B_\textrm{est} \approx B$,
 where $A_\textrm{est}$ and $B_\textrm{est}$ \textit{are} compatible.  In quantum mechanics, 
 physical quantities  can always be \blu represented by Hermitian operators  (in some Hilbert space), 
 and the above statements imply $[\hat A, \hat B] \neq 0$ but $[\hat A_\textrm{est}, \hat B_\textrm{est}] = 0$. The key question then concerns the \gold \textit{quality}  of \gold these \blk 
 jointly-measurable  estimators: \blk how close are they to the original observables of interest?  
  A natural 
measure of inaccuracy is 
the root mean square error, \blu $\epsilon(A_{\rm est}):= 
\blu \langle (\hat{A}_{\rm est} - \hat{A})^2\rangle^{1/2}$ \blk \cite{vantrees,ozawa1}. 
This measure vanishes for a perfect estimate, \blu $\hat{A}_{\rm est}=\hat{A}$, \blk and more generally quantifies the degree to which the  estimator \blk 
deviates from the quantity being estimated \cite{note}.  

Arthurs and Kelly \blu showed \blk in 1965  that $\epsilon(Q_{\rm est})   \epsilon(P_{\rm est}) \geq \hbar/2$  for any joint estimates of position and momentum   \blu that \blk are \textit{globally unbiased},  i.e., for which the average values of the estimators are equal to the average values of the corresponding observables for every  state  
\cite{ak}. 
\blu This \blk relation  was subsequently generalized to the complementarity relation \cite{ozawa1,ag,ishi}
\begin{equation} \label{ak}
\epsilon(A_{\rm est})\, \epsilon(B_{\rm est}) \geq  \frac{c}{2} , \blk
\end{equation}
\blu valid \blk for globally-unbiased joint estimates of \blu any two \blk observables $A$ and $B$.  Here $ c \blk :=|\langle [\hat{A},\hat{B}]\rangle|$, and is nonzero only if $A$ and $B$ are incompatible. 
\blu   Thus, the more accurately a  globally-unbiased  joint measurement  can estimate  observable $A$, the less accurately it can estimate an incompatible observable $B$, and vice versa.  \blk  Other complementarity relations---also called joint-measurement uncertainty relations---have been given, \blu  using alternative measures of inaccuracy, \blk but are  again  only valid for restricted classes of measurements \blu \cite{woot, mart, jaeger, apple, trif,busch, watanabe}.  \blk

The above-mentioned complementarity relations not only lack universal validity  but, most importantly,  
\grn are inapplicable \blk  in cases of considerable physical interest.  For example, Einstein, Podolsky and Rosen (EPR) based their argument for the incompleteness of quantum theory on the case of two particles having  highly correlated  positions  and momenta \cite{epr}.  In this case, one can make  highly \blk accurate joint estimates of the position and momentum of the first particle via,   for example,  a position measurement on the first particle and a momentum measurement on the second. Thus, the Arthurs-Kelly relation (\ref{ak}) clearly fails in the EPR scenario.

\begin{figure*}[!ht]
\centering
\includegraphics[width=18cm]{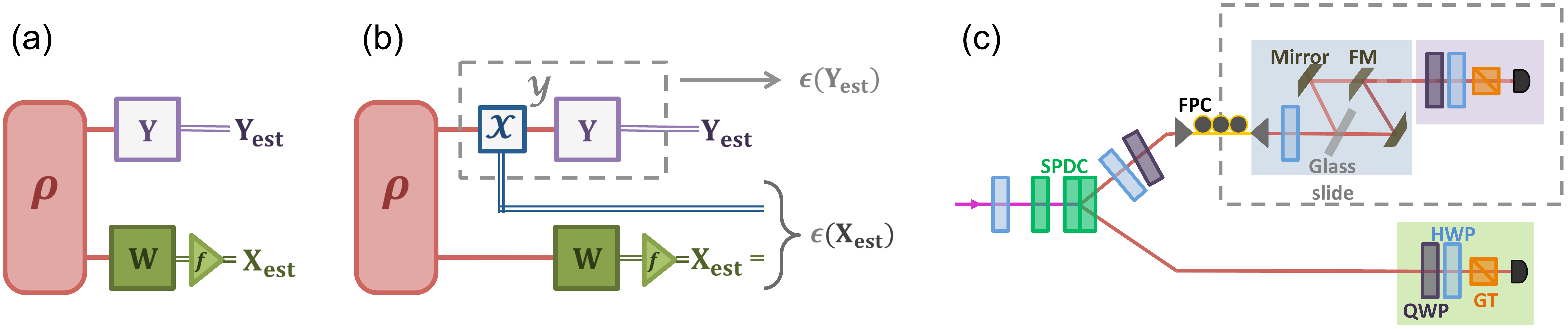}
\caption{(Color online). Concept and implementation of the experiment.   (a) A source generates entangled pairs of optical qubits.  Ideally, a joint measurement, comprising a $Y$-polarisation measurement on the first qubit and a polarisation measurement on the second qubit in some direction $W$, is used to simultaneously estimate the $X$ and $Y$ polarisations of the first qubit, \grn via 
\blk $X_{\rm est}=f(W)$ and $Y_{\rm est}=Y$, \grn giving \blk  
 $\epsilon(Y_{\rm est})=0$. (b) To determine $\epsilon(X_{\rm est})$, a measurement  
${\cal X }$ probing $X$ can be 
made \blk prior to the $Y$ measurement. 
The \blk inaccuracy, $\epsilon(X_{\rm est})$, of any estimate $X_{\rm est}=f(W)$  is obtained from the joint statistics of $X_{\rm est}$ and the outcome of  
${\cal X}$, without any knowledge of 
$W$ or $f$ or $\rho$ (as indicated by their shaded boxes). If   
${\cal X}$ \grn were \blk  arbitrarily weak, we would still have $\epsilon(Y_{\rm est})\approx 0$, but in practice  it is only semiweak. Nevertheless \grn with \blk the measurements ${\cal X }$  and $Y$ fully characterized (as indicated by their unshaded boxes) we can calculate $\epsilon(Y_{\rm est})$ from the total operation \grn ${\cal Y}$ \blk 
that \blk yields $Y_{\rm est}$. 
   (c) The implementation of \blk  ${\cal X }$ uses a glass slide as a  polarisation-dependent  beam splitter to make a semiweak measurement of $X$.  A \blk flip mirror (FM)  selects between the reflected and transmitted beams to enable the subsequent measurement of $Y$. Optical elements are coded by colour,  with representative examples labeled. \blk  QWP and HWP denote a quarter wave and half wave plate, respectively; FPC a fibre-polarisation controller;  GT a Glan-Taylor prism, and SPDC the spontaneous parametric down-converter.}
\label{fig1}
\end{figure*}

Indeed, it is only relatively recently that universally valid complementarity relations have been given, holding for arbitrary joint estimates of two observables in any measurement scenario.  In particular, following related work  by Ozawa  on  measurement-disturbance  relations \cite{ozawadist}, Hall showed that  the inaccuracies of any estimates generated by  a joint measurement of observables $A$ and $B$ must satisfy \cite{hall}
\begin{equation} \label{hall}
\epsilon(A_{\rm est}) \,\epsilon(B_{\rm est}) + \epsilon(A_{\rm est})\, \Delta B_{\rm est} + \Delta A_{\rm est} \,\epsilon(B_{\rm est}) \geq  \frac{c}{2} \blk ,
\end{equation}
where  $\Delta G := (\langle \hat{G}^2\rangle-\langle \hat{G}\rangle^2)^{1/2}$ is the spread in quantity $G$. \gold Thus, unlike Eq.~(\ref{ak}), the inaccuracies can both be arbitrarily small, provided the spreads of the estimates are sufficiently large.  This is a consequence of not imposing global unbiasedness: \grn for a given state \blk one can now tailor joint estimation schemes to the specific state of interest (where these schemes may perform quite poorly for other states), so as to reduce \blk the inaccuracies.  However, the {\it spreads} will be correspondingly increased.  The above EPR scenario provides such a tailored estimation scheme, for position and momentum, which saturates Eq.~(\ref{hall}) \cite{hall}. \blk  Hall also determined the form of the {\it optimal} estimates of $A$ and $B$ in any given measurement scenario, i.e., those functions of the measurement \blu data  that yield estimators $A_\textrm{est}$ and $B_\textrm{est}$ \blk with the smallest possible inaccuracies $\epsilon(A_{\rm est})$ and $\epsilon(B_{\rm est})$ \cite{hall,eur}.  
 
Ozawa has given a formally similar relation \cite{ozawa}
\begin{equation} \label{ozawa}
\epsilon(A_{\rm est})\, \epsilon(B_{\rm est}) + \epsilon(A_{\rm est})\, \Delta B+ \Delta A \,\epsilon(B_{\rm est}) \geq  \frac{c}{2} \blk .
\end{equation}
This differs from Eq.~(\ref{hall}) in  depending on the spreads \blu $\Delta A$ and $\Delta B$  of \grn the observables, \blk rather than on the actual measured spreads $\Delta A_\textrm{est}$ and $\Delta B_\textrm{est}$, but similarly implies that the inaccuracies cannot both be zero for incompatible observables. \blk The Hall relation \blu implies (and hence is stronger than) \blk the Ozawa relation for the \blu particular \blk case of optimal  estimators, \grn for which \blk $\Delta A_{\rm est} \leq \Delta A$ \cite{hall,eur}.

Here we provide the first experimental test of universal complementarity relations, 
in an EPR-type scenario for which the Arthurs-Kelly relation (\ref{ak}) is shown to be violated. Moreover, a  new universal complementarity relation, which is stronger than both the Hall and Ozawa relations for optimal estimates, is proved here and similarly   tested. \blk  The measurement inaccuracies are determined  by generalizing \grn the Lund-Wiseman \blk method \cite{lund} (recently implemented in the context of measurement--disturbance relations \cite{rozema}),  by \blu using the \blk contextual value formalism \cite{dress1} to allow the replacement of a weak measurement interaction by one of arbitrary strength. \blk

{\it  Stronger \blk universal complementarity relation.}---In addition to  experimentally investigating the complementarity relations (\ref{ak})--(\ref{ozawa}), we also tested the relation
\begin{equation} \label{hall2}
\epsilon(A_{\rm est})\, \frac{\Delta B_{\rm est}+\Delta  B}{2} + \epsilon(B_{\rm est})\,\frac{\Delta A_{\rm est}+\Delta  A}{2}\geq  \frac{c}{2} \blk ,
\end{equation}
which we have derived  (see  the Supplemental Material \cite{supp}). 
Unlike Eqs.~(\ref{ak})--(\ref{ozawa}), this relation does not involve an $\epsilon(A_{\rm est})\epsilon(B_{\rm est})$ term, and so directly quantifies the tradeoff between inaccuracy and spread. \blk It similarly implies that $\epsilon(A_{\rm est})$ and $\epsilon(B_{\rm est})$ cannot both vanish for incompatible $A$ and $B$.

The complementarity relation (\ref{hall2}) is clearly stronger than a simple average of Eqs.~(\ref{hall}) and (\ref{ozawa}), \blu and strictly implies each of the latter in the case where optimal estimates are made from the measurement data \cite{supp}.  It is also stronger, in the sense of \gold lying closer to the bound, \blu  for all  estimators \blk considered in our experiment.  
\blk

{\it Experimental setup.}---To provide a discrete analogy  of  the EPR scenario discussed in the introduction, we generated pairs of optical polarisation qubits \blu (see Fig.~1) \blk having high fidelity with the entangled state
\begin{equation} \label{psi}
|\psi\rangle = \cos\gamma |HV\rangle - \sin \gamma|VH\rangle .
\end{equation}
Here $H$ and $V$ refer to  horizontal and vertical polarisation, which are assigned as eigenstates of the $\hat{Z}$  polarisation \blk operator \cite{WisMil10}.  Observables $A$ and $B$ in the complementarity relations (\ref{ak})-(\ref{hall2}) are chosen to be the $X$ [diagonal (anti-diagonal)] and $Y$ [right-(left-)circular]  polarisations of the first qubit. \blk

The state $|\psi\rangle$ \blk is maximally entangled for $\gamma=\pi/4$,  allowing perfectly accurate joint estimates  of   $X$  and $Y$ \blk (via  measuring \blk $Y$   directly, \blk and the $X$ polarisation of  the perfectly  anticorrelated \blk second qubit), analogous to the EPR example.  However, 
the lower bound of the complementarity relations, $c/2=|\langle \hat{Z}\rangle|=|\cos 2\gamma|$, trivially vanishes.
 Conversely,  the choice $\gamma=0$ (or $\pi/2$) maximises  $c$, \blk but \blk corresponds to a factorisable state from which  no information about the 
 the first qubit can be gained via a measurement on the second.  Hence,  the \blk intermediate choice $\gamma=\pi/8$ was made for our experiment.  Quantum state tomography \cite{tom} on the state generated by  a  spontaneous parametric down-conversion source gave a fidelity of $97.4\pm 0.3\%$  with the desired entangled state (see the Supplemental Material \cite{supp}). \blk

\red We demonstrate how to experimentally use semiweak measurements, in conjunction with strong measurements, to operationally determine the inaccuracy $\epsilon$. \blk The aim of the experiment is to jointly estimate the $X$  and $Y$  polarisations of the first qubit.  
\blk  One could directly \blk measure $Y_\textrm{est}=Y$ on the first qubit, and   estimate \blk $X_\textrm{est}=f(W)$ by applying some function to the  outcome of a polarisation observable \blk $W$ measured on  the second \blk qubit  [see Fig.~1(a)]. \blk  However, to \blk allow the corresponding inaccuracies to  be determined, the first qubit was also subjected to an initial semiweak measurement process ${\cal X }$ \cite{pryde05}  [see Fig.~1(b)], \blk implemented using the polarisation dependence of Fresnel reflection from a  glass  slide \cite{dress2}  oriented at an angle $\beta =75^\circ$ [see Fig.~1(c)]. \red While weak measurements are measurements that introduce negligible disturbance to a system while providing an infinitesimal amount of information, semiweak measurements provide small but finite information about the system, with small but non-negligible back action. \blk

The slide reflects (transmits) \blk horizontal and vertical polarisations with probabilities $r_H$ and $r_V$ \blu ($t_H=1-r_H$ and $t_V=1-r_V$), found by direct measurement to be \blk $r_H=0.1244\pm 0.0001$  and $r_V=0.4645\pm 0.0011$.  Wave plates before and after the slide were adjusted to rotate the basis of the semiweak measurement from \grn ${Z}$ \blk to $X$. 
The strength of the measurement $\cal X$ is quantified by  $\kappa:=1-\sqrt{r_Hr_V}-\sqrt{t_Ht_V} \in [0,1]$. 
The above values of $r_H$ and $r_V$ give $\kappa=0.0749\pm 0.008$,  indicating that ${\cal X}$ is semiweak. 
 \blk

\blu
{\it Measuring the inaccuracies.}---The outcome of the semiweak measurement, $m=r$ or $t$, occurs with some probability $p(m)$, and probes the $X$ polarisation of the first qubit. It has a  unique \blu associated numerical value, $\xi_m$, called the {\it contextual value} of $X$ \cite{dress1}, defined by the property $\sum_m \xi_m p(m) \equiv \langle \hat{X}\rangle$, thus generalising the notion of eigenvalues to semiweak measurements. The contextual values are simple functions of $r_H$ and $r_V$ (see the Supplemental Material \cite{supp} and Ref. \cite{dress2}). As shown in the Supplemental material \cite{supp}, they allow \blk the inaccuracy of any  estimator \blk of the form $X_{\rm est}=f(W)$ to be \blu experimentally \blk determined \blu via the formula \blk
\begin{equation} \label{xest}
\epsilon(X_{\rm est})^2 = \blu \frac{1}{2} \blk \sum_{x,m,y,w} [x-f(w)]^2\, \blu (1+x\xi_{m}) \blk \,p(m,y,w)    .
\end{equation}
Here $p(m,y,w)$ is the  measured joint  probability of  outcomes $m$, $y$ and $w$ for the binary measurements $\cal X$, $Y$ and $W$, respectively, and $x=\pm1$. 
 Equation~(\ref{xest}) 
\blu reduces to \blk Eq.~(16) of Ref.~\cite{lund} in the weak limit,  
$\kappa \to 0$. \blk

 It would be possible to similarly determine the inaccuracy of the  estimator \blk $Y_{\rm est}$ by rotating the basis of the semiweak measurement ${\cal X}$ from $X$ to $Y$.  However, this inaccuracy can instead \blk be inferred by taking into account the effect of ${\cal X }$ on the first qubit  [Fig.~1(b)].  In particular, \blu as shown in the Supplemental Material \cite{supp},  the measurement of $Y$ subsequent to 
 ${\cal X }$ is  equivalent to a generalized measurement ${\cal Y}$  on the initial state, described by \blu a  positive-operator-valued \blk measure $\{\hat{\Upsilon}_\pm\}$, \blu with \blk $\hat{\Upsilon}_\pm=\frac{1}{2} \pm\frac{1}{2}(1-\kappa) \hat{Y}$.  
 The \blk inaccuracy of the  estimator \blk $Y_{\rm est}$  
  corresponding to $\cal{Y}$  follows as \blk \gold $\epsilon(Y_{\rm est})^2  =2\kappa  = 0.15 \pm 0.02$\blk ,
irrespective of the input state (see the Supplemental Material \cite{supp}).  Note the inaccuracy vanishes in the weak limit $\kappa   \rightarrow   0$. 

{\it Experimental results.}---The joint probability distribution $p(m,y,w)$ was measured for several choices of the polarisation observable $W$, corresponding to the Bloch sphere angles $\theta=90^\circ$ and $\phi=135^\circ,157.5^\circ,180^\circ,202.5^\circ$,   and  $225^\circ$.  \grn For state $|\psi\rangle$ in (\ref{psi}), \blk $\theta=90^\circ$ and $\phi=180^\circ$ correspond to the observable of the second qubit that is most strongly correlated with the $X$ polarisation of the first, 
 in the sense of maximising $\langle \hat{X}\otimes \hat{W}\rangle$. \gold Data were collected for 30 s per setting, with a total measured flux of 2000 coincidence counts per second (see the Supplemental Material \cite{supp}). \blk

Two types of estimates for $X$, based on the outcome $w=\pm 1$ of the $W$ measurement, were considered.  The first type corresponded to simply estimating $X$ to be $w$, i.e.,  $X_{\rm est}^{\rm simple}=W$,  which is \blu the best possible \blk if $\hat{X}_{\rm est}$ is constrained to have 
the same eigenvalues as $\hat{X}$. \blk 
The second type of estimate exploited the tomographically determined state of the source, $\hat{\rho}$, to make the {\it optimal} estimate, $X_{\rm est}^{\rm opt}=f_{\rm opt}(W)$, \blu corresponding to the smallest possible inaccuracy, \blk with  \cite{hall}
$$ f_{\rm opt}(w) :=  \an{\hat{X}\otimes  \ro{\hat{\mathbbm{1}}+w\hat{W}}}_{\hat{\rho}} \div \an{\hat{\mathbbm{1}}\otimes  \ro{\hat{\mathbbm{1}}+w\hat{W}}}_{\hat{\rho}}.$$
For   example, for $W$  defined by \blk  
$\theta=90^\circ$ and $\phi=180^\circ$, \blk  $f_{\rm opt}(1)=0.630$ and $f_{\rm opt}(-1)=-0.643$.  

 The inaccuracies were determined via Eq.~(\ref{xest}), and are plotted in Fig.~2.  It  is \blk also verified  in Fig.~2  that the inaccuracy and spread of $X_{\rm est}^{\rm opt}$  satisfy  Hall's \blk inaccuracy--dispersion relation
\begin{equation} \label{dis}
\epsilon(X_{\rm est}^{\rm opt})^2 + (\Delta X_{\rm est}^{\rm opt})^2 = (\Delta X)^2
\end{equation}
for optimal estimates  \cite{hall}. 

\begin{figure}[!ht]
\centering
\includegraphics[width=0.45\textwidth]{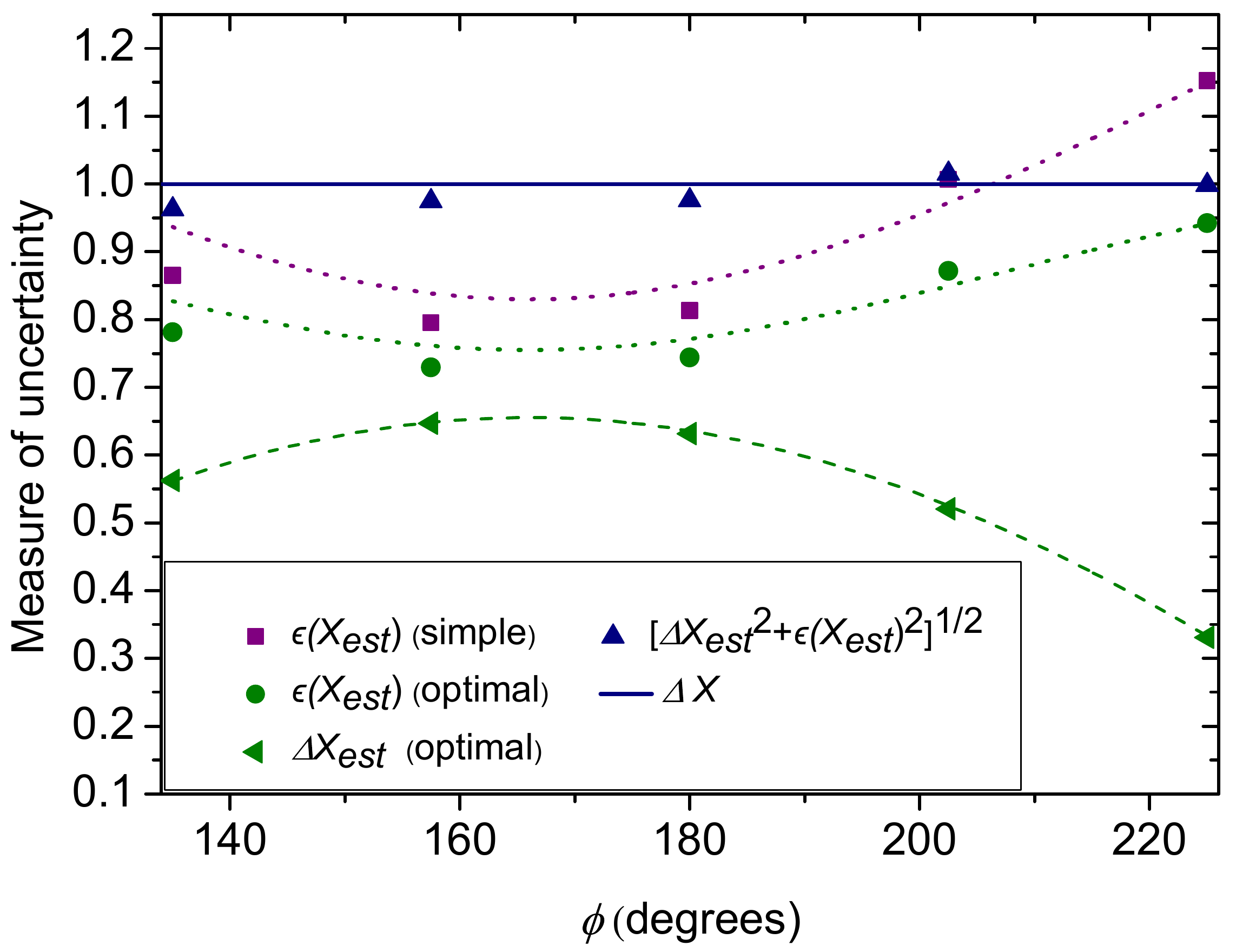}
\caption{(Color online). Inaccuracies of simple and optimal estimates of $X$ from a measurement of $W$. Theoretical curves are given for the inaccuracies $\epsilon(X_{\rm est}^{\rm simple})$ and  $\epsilon(X_{\rm est}^{\rm opt})$ (upper and lower dotted lines), and for the spread $\Delta X_{\rm est}^{\rm opt}$ (dashed line).  The solid line represents the tomographically determined value of $\Delta X$, with the closely adjacent data points corresponding to the square root of the left hand side of Eq.~(\ref{dis}).  Error bars not shown are smaller than the size of the markers. }
\label{fig2}
\end{figure}

The  spreads $\Delta X_{\rm est}$ and $\Delta Y_{\rm est}$ were calculated directly from the  measured distribution $p(m,y,w)$, \blk  while we obtained \blk  $\Delta X=0.998\pm0.002$, $\Delta Y=0.9998\pm  0.0001$, and 
 $c/2=|\langle \hat{Z}\rangle|=0.711\pm 0.004$ \blk from $\hat{\rho}$.  
\blu Combined with the  above \blu inaccuracy data, \blu this \blk allowed complementarity relations (\ref{ak})--(\ref{hall2}) to be tested for the two types of estimate of $X$. The universal complementarity relations (\ref{hall})--(\ref{hall2}) were validated and the Arthurs-Kelly relation  (\ref{ak}) \blk   strongly \blk violated (Fig.~3). The latter violation was expected for EPR scenarios, as discussed in the introduction.  Fig.~3(b) also verifies that,  for optimal estimates, \blk the new complementarity relation (\ref{hall2}) is stronger than relations (\ref{hall}) and (\ref{ozawa}),  \blk as expected.

\begin{figure}[!ht]
\centering
\includegraphics[width=0.48\textwidth]{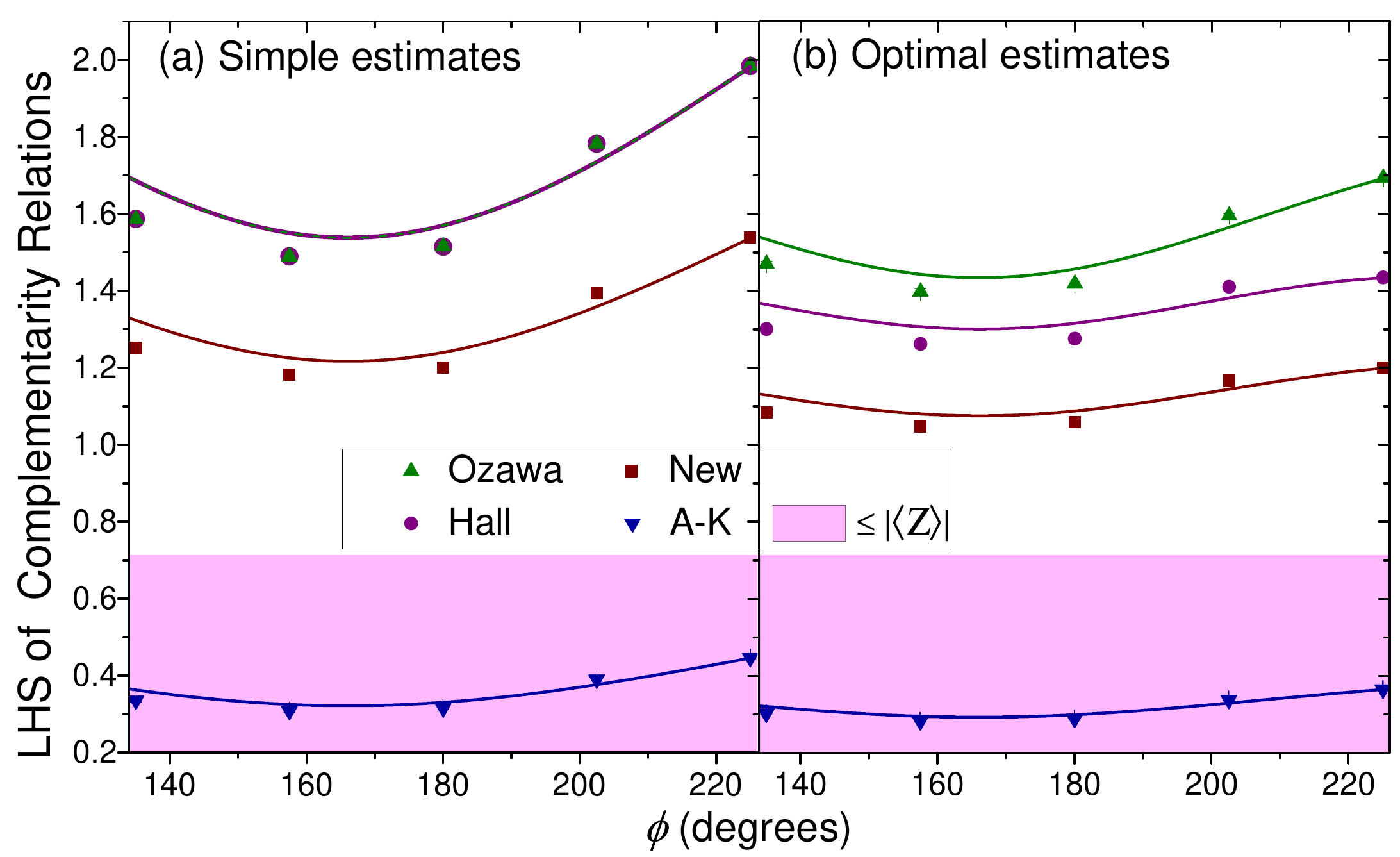}
\caption{(Color online). Experimental test of complementarity relations.   Subfigures (a) and (b) \blk correspond to the simple and the optimal estimates of $X$ described in the text. Each  subfigure \blk shows, in descending order, the left hand sides of the \red Ozawa and Hall relations (\ref{ozawa}) and (\ref{hall}) \blk [indistinguishable in  (a)]; the new complementarity relation (\ref{hall2}); and the left hand side of the Arthurs-Kelly relation (\ref{ak}).  The   pink  shading indicates the region corresponding to violation of any of these relations.  Error bars not shown are smaller than the size of the markers.   Solid curves are theoretical predictions.  }
\label{fig3}
\end{figure}

{\it Discussion}---Universal complementarity relations \blu should \blk hold for any joint measurement of two observables. We have provided the first experimental \blu investigation \blk of such relations (Fig.~3). This was done in an EPR-type scenario, for which the Arthurs-Kelly relation (\ref{ak}) is violated.  We have, furthermore, derived and verified a new universal complementarity relation, Eq.~(\ref{hall2}), that is significantly stronger than the previously obtained relations (\ref{hall}) and (\ref{ozawa}) for the case of optimal  estimators. We \blk have also verified the inaccuracy--dispersion relation (\ref{dis}) for optimal  estimators.  \blk

It would be of interest to  also test  the universal complementarity relations for the case of 
continuously valued quadrature observables   $\hat{Q}=\hat{a}+\hat{a}^\dagger$ and $\hat{P}=(\hat{a}-\hat{a}^\dagger)/i$.  This would allow the relations to be  explored in a  context more akin to the original scenario envisaged by \grn EPR \blk \cite{epr}, with \blk the advantage  
 of a saturable and fixed lower bound~\cite{hall}. 

Our work represents an important advance in the quantitative understanding and experimental verification of complementarity, arguably the most important foundational principle of quantum mechanics.  This principle underlies many aspects of quantum information technology, ranging from entanglement verification \cite{buscemi} to quantum dense coding \cite{dense} to the security of quantum cryptography \cite{crypt}, and our work could have implications in all these areas. 

\acknowledgments {\it Acknowledgments}---MMW thanks Dylan Saunders, Adam Bennet, Allen Boston and Sacha Kocsis for helpful contributions. MJWH thanks Cyril Branciard for pointing out the difference between the joint uncertainty relations in Refs.~\cite{hall} and \cite{ozawa}, and for stimulating discussions.  \blu {We are grateful to Yutaka Shikano and Masanao Ozawa for drawing our attention to relevant literature.  \blk This research was supported by the ARC Centre of Excellence CE110001027.

\onecolumngrid
\newpage
\begin{center}
{\large \textbf{Supplemental Material}}
\end{center}

\twocolumngrid  
  
\section{ Proof of Eq.~(4)} 

Any two joint estimates of observables $A$ and $B$ may be represented by two commuting operators $\hat{A}_{\rm est}$ and $\hat{B}_{\rm est}$ on a suitable Hilbert space (via a Naimark extension if necessary) \cite{hall, ozawa}.  It follows that $2 [\hat{A},\hat{B}]= [\hat{A}-\hat{A}_{\rm est},\hat{B}+\hat{B}_{\rm est}] + [\hat{A}+\hat{A}_{\rm est},\hat{B}-\hat{B}_{\rm est}]$, and hence, using the triangle inequality, that
\begin{eqnarray*}
2|\langle [\hat{A},\hat{B}]\rangle | &=& |\langle [\hat{A}-\hat{A}_{\rm est},\hat{B}] +[\hat{A}-\hat{A}_{\rm est},\hat{B}_{\rm est}]\\
&~& + [\hat{A},\hat{B}-\hat{B}_{\rm est}]+ [\hat{A}_{\rm est},\hat{B}-\hat{B}_{\rm est}]\rangle |\\
&\leq& |\langle [\hat{A}-\hat{A}_{\rm est},\hat{B}]\rangle| + |\langle [\hat{A}-\hat{A}_{\rm est},\hat{B}_{\rm est}]\rangle|\\
&~& + |\langle[\hat{A},\hat{B}-\hat{B}_{\rm est}]\rangle|+ |\langle[\hat{A}_{\rm est},\hat{B}-\hat{B}_{\rm est}]\rangle | .
\end{eqnarray*}
Applying the Schwarz inequality  (which lies behind all \blu such \blk relations), \blk $ |\langle [\hat{R},\hat{S}]\rangle|\leq 2 \sqrt{\langle (\hat{R}-r)^2\rangle \, \langle (\hat{S}-s)^2\rangle}$, with $r$ and $s$  suitably  chosen from $0$, $\langle \hat{R}\rangle$ and $\langle \hat{S}\rangle$, leads to Eq.~(4) as desired.

Eq.~(4) is stronger than complementarity relations (2) and (3), both in the sense of implication and of having a smaller left hand side, for the case that  optimal estimates $A_{\rm opt}$ and $B_{\rm opt}$ are made [corresponding to those functions of the measurement \blu data \blk that yield the smallest possible inaccuracies $\epsilon(A_{\rm est})$ and $\epsilon(B_{\rm est})]$. This follows from the  inaccuracy-dispersion relations
$(\Delta A)^2=(\Delta A_{\rm opt})^2 +\epsilon(A_{\rm opt})^2$ and
$(\Delta B)^2=(\Delta B_{\rm opt})^2 +\epsilon(B_{\rm opt})^2$ \blk
for optimal estimates \cite{hall}. In particular, $\Delta A_{\rm opt}\leq\Delta A$ and $\Delta B_{\rm opt}\leq\Delta B$, and hence Eq.~(4) implies Eq.~(3) for such estimates.  Further, defining $\alpha:=\epsilon(A_{\rm opt})/\Delta A$ and $\beta:=\epsilon(B_{\rm opt})/\Delta B$, then $0\leq \alpha,\beta\leq 1$, and the  difference of the left hand sides of relations (2) and (4) is $\epsilon(A_{\rm opt})\Delta B\,h(\beta)+\Delta A\,\epsilon(B_{\rm opt})h(\alpha)$, where $h(x):=(1/2)[\sqrt{1-x^2}-(1-x)]$ is never less than zero.  Hence Eq.~(4) also implies Eq.~(2) for optimal estimates, as claimed. 

\blu
It is of interest to note that  if $\eta(B):=\langle (\hat{B}'-\hat{B})^2\rangle^{1/2}$ denotes the disturbance to an observable $B$, caused by making an estimate $A_{\rm est}$ of $A$ that changes $B$ to $B'$ \cite{ozawadist,ozawaannphys}, then one may similarly derive a measurement-disturbance uncertainty relation,
\[ \epsilon(A_{\rm est})\, \frac{\Delta B +\Delta  B'}{2} + \eta(B)\,\frac{\Delta A_{\rm est}+\Delta  A}{2}\geq  \frac{c}{2} \blk , \]
as will be discussed elsewhere.  \red We note also that new joint-measurement and measurement-disturbance relations have recently been given by Branciard \cite{branc}.
\blk

\section{Photon source} 

A  120~mW, linearly polarised continuous wave  404~nm laser diode was used to pump a pair of  0.5~mm thick Bismuth Borate    (BiBO) crystals.  The crystal pair is oriented such that the optical axes of the pair are crossed in what is known as a `sandwich', and an additional BiBO crystal is inserted into the pump to precompensate the temporal walkoff between H and V polarisations in the downconversion crystal \cite{source1,source2}.  When pumped with  diagonally polarised light, spontaneous parametric down conversion leads to photon pairs at 808~nm in the maximally entangled $|\Phi^{+}\rangle $ state. \red By adjusting a half wave plate inserted into the pump beam before the crystal (Fig. 1(c) of the main paper), we can tune $\gamma$ (Eq.(5)). To convert $|\Phi^{+}\rangle$ to $|\psi\rangle$, we must also change the $Z$-basis correlation and the phase of the superposition, which we do by applying an $X$ and a $Z$ operation to one of the qubits. Experimentally, we implement these operations on the first qubit using wave plates immediately following the SPDC source, as shown in Fig. 1(c) of the main paper. \blk Counting in coincidence with a 3~ns window and  3~nm FWHM interference filters,  photon pair rates of  $\sim 2000$~s$^{-1}$ were observed. Quantum tomography was used to reconstruct the state produced by the source, as shown in Fig.~4.

\begin{figure}[!ht]
\centering
\includegraphics[width=0.50\textwidth]{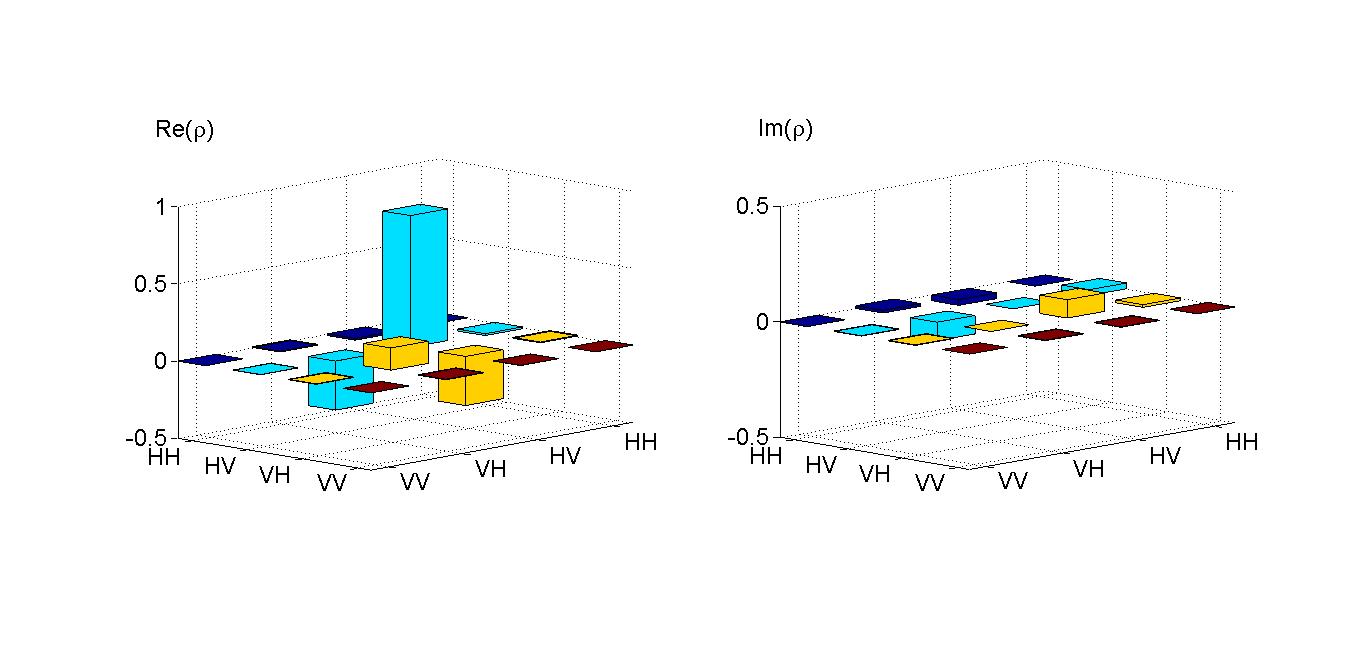}
\caption{(Color online). Tomographic reconstruction of the two-qubit state generated by the source, showing the real and imaginary parts of the density operator in the H-V basis.  The state has a fidelity of $97.4\pm0.3\%$ with the desired entangled state in Eq.~(4) of the main text (with $\gamma=\pi/8$). }
\end{figure}

\blu

\section{Data and and error analysis}

\begin{table*}[ht]
\centering
\caption{The measured joint probability $p(m,y,w)$, for various orientations of polarisation observable $W$. Here
$m$, $y$, and $w=\pm 1$ are the binary measurement outcomes for measurements ${\cal X}$, $Y$ and $W$, respectively (see Fig~1(b) of main text).  The Bloch sphere angles parameterising the orientation of $W$ are $\theta=90^\circ$ and $\phi=135^\circ$, $157.5^\circ$, $180^\circ$, $202.5^\circ$ and $225^\circ$. }

\begin{tabular*}{\textwidth}{c @{\extracolsep{\fill}} ||c|c|c|c|c|c}

\hline $p(m,y,w)$    & $\phi=135^\circ $ & $\phi =157.5^\circ$ & $\phi= 180^\circ  $ & $\phi= 202.5^\circ$ & $\phi= 225^\circ  $  \\ 
\hline
\hline $p(1,1,1)$    & $ 0.195(2) $ & $ 0.234(2)  $ & $ 0.282(2)  $ & $ 0.323(2)  $ & $ 0.327(2)  $ \\ 
\hline $p(1,1,-1)$   & $ 0.208(2) $ & $ 0.174(2)  $ & $ 0.112(2)  $ & $ 0.084(1)  $ & $ 0.0712(9) $ \\ 
\hline $p(1,-1,1)$   & $ 0.228(2) $ & $ 0.208(2)  $ & $ 0.157(2)  $ & $ 0.112(2)  $ & $ 0.086(1)  $ \\ 
\hline $p(1,-1,-1)$  & $ 0.088(1) $ & $ 0.116(2)  $ & $ 0.162(2)  $ & $ 0.206(2)  $ & $ 0.234(2)  $ \\ 
\hline $p(-1,1,1)$   & $ 0.1951(4)$ & $ 0.0193(5) $ & $ 0.0363(7) $ & $ 0.0541(8) $ & $ 0.0680(9) $ \\ 
\hline $p(-1,1,-1)$  & $ 0.208(1) $ & $ 0.0732(9) $ & $ 0.0607(9) $ & $ 0.0405(7) $ & $ 0.0300(6) $ \\ 
\hline $p(-1,-1,1)$  & $ 0.228(1) $ & $ 0.0642(9) $ & $ 0.0614(9) $ & $ 0.0540(8) $ & $ 0.0549(8) $ \\ 
\hline $p(-1,-1,-1)$ & $ 0.088(2) $ & $ 0.112(2)  $ & $ 0.129(2)  $ & $ 0.127(2)  $ & $ 0.129(2)  $ \\ 
\hline 

\end{tabular*} 

\label{jointprob}
\end{table*}

 The measured relative frequencies for the outcomes of ${\cal X}$, $Y$ and $W$ in Fig~1(b) of the main text are shown in Table~I below. \blk Uncertainties predominantly arise from Poissonian counting statistics due to the random SPDC generation times, and slow, small thermal drifts in the fibre coupling leading to changes in $r_H$ and $r_V$ during the data collection. The latter effect was characterized by measuring these quantites before and after the data collection, and observing the mean and spread of the values. Uncertainties in directly measured experimental values were combined using standard error propagation formulae to generate uncertainties in values calculated from the data. The small deviations between experiment and theory are attributed to a combination of inexact wave plate settings and other uncompensated small thermal drifts (e.g.\ fibre birefringence) in the experiment.

\blu

\section{Contextual values for the semiweak measurement ${\cal X}$}

As noted in the main text, contextual values for the outcomes of semiweak measurements were introduced by Dressel {\it et al.}, and generalise the notion of eigenvalues for projective measurements \cite{dress1,dress2}.  As was further noted, the contextual values $\xi_r$ and $\xi_t$ of the semiweak measurement ${\cal X}$ in Fig.~1(b), corresponding to reflection and transmission by the glass slide in Fig.~1(c), are defined by the property $\sum_{m=r,t} \xi_m p(m)=\langle \hat{X}\rangle$, where $p(r)$ and $p(t)$ denote the probability of reflection and transmission, respectively. 

To determine $\xi_m$, note that the operation of the glass slide, combined with wave plates before and after it to rotate the measurement basis from the $Z$ polarisation to the $X$ polarisation, corresponds to a (semiweak) polarisation-dependent beamsplitter.  As per Dressel {\it et al.} \cite{dress2} (who analysed the case of $Z$ polarisation),  the action of ${\cal X}$ on any input qubit state $\hat{\tau}$ is therefore described by the completely-positive trace-preserving map 
\beq \label{mh} \mu(\hat{\tau}) := \hat{M}_r \hat{\tau}  \hat{M}_r + \hat{M}_t \hat{\tau}  \hat{M}_t ,\eeq
where the corresponding measurement operators $\hat{M}_r$ and $\hat{M}_t$ are Hermitian with  $\hat{M}_r^2 = r_H\hat{X}_+ +r_V\hat{X}_-=\hat{\mathbbm{1}}-\hat{M}_t^2$, with $\hat{X}_\pm$ denoting $\frac{1}{2}(\mathbbm{1}\pm \hat{X})$  and $\hat{\mathbbm{1}}$  the unit operator. The probabilities of reflection and transmission follow as $p(m)={\rm Tr}[\hat{\tau} \hat{M}_m^2]$.  The equation $\sum_{m} \xi_m p(m)=\langle \hat{X}\rangle$ may then be solved for the contextual values $\xi_r$ and $\xi_t$, to give \cite{dress2}
\beq \xi_r = \frac{2-r_H-r_V}{r_H-r_V},~~~\xi_t=-\frac{r_H+r_V}{r_H-r_V} .  \eeq

Note that in the `strong' limit $|r_H-r_V|=1$, corresponding to perfect discrimination of the eigenstates of $\hat{X}$ by the measurement ${\cal X}$, the contextual values are equal to the associated eigenvalues $\pm 1$ of $\hat{X}$.  Conversely, in the `weak' limit $r_H\to r_V$, corresponding to polarisation-independent operation, with no information about $X$ able to be gained from the measurement ${\cal X}$, the contextual values become unbounded.  For the measured values of $r_H$ and $r_V$ in our experiment (given in the main text), the associated contextual values follow as $\xi_r=-1.73$ and $\xi_t=4.15$.

\section{Determining inaccuracies from semiweak measurements}

\blk
To test the validity \blu or otherwise \blk of complementarity relations (1)--(4), it \blu is \blk necessary to experimentally determine the inaccuracies $\epsilon(A_{\rm est})$ and $\epsilon(B_{\rm est})$.  This problem has been previously  discussed  in context  of  measurement--disturbance uncertainty relations, which also involve the inaccuracy of an estimate.  Ozawa proposed an approach in which  $\epsilon(A_{\rm est})$ is determined from the statistics of (separate) measurements of $A$ and $A_{\rm est}$, on each of three different states of the system (defined by $A$ and the state of interest) \cite{ozawaannphys}.  Lund and Wiseman  proposed an approach not requiring preparation of additional states, in which a weak measurement of $A$ is made prior to the measurement of $A_{\rm est}$ \cite{lund}, allowing $\epsilon(A_{\rm est})$ to be determined from the corresponding  weak-valued joint probability distribution.  These two  proposals have been recently applied to the verification of measurement--disturbance uncertainty relations in  Refs.~\cite{ozawanat} and \cite{rozema} respectively. 

Our approach is a   generalisation  of the Lund-Wiseman proposal, in which the weak measurement interaction is replaced by one of arbitrary strength.  To introduce the basic concept, 
consider two quantum observables $K$ and $L$, \blu represented by Hermitian operators $\hat{K}$ and $\hat{L}$ \blk having respective eigenvalue decompositions $\hat{K}=\sum_k \kappa_k \hat{K}_k$ and $\hat{L}=\sum_l \lambda_l \hat{L}_l$.  The corresponding \textit{Margenau-Hill} joint quasiprobability distribution is \blu then \blk defined by  \cite{mh}
\[ p_{\rm MH}(k,l):= \langle \hat{K}_k\hat{L}_l +\hat{L}_l\hat{K}_k\rangle/2 . \] 
It can take  negative values for non-commuting observables, but is normalised and has marginal distributions corresponding to the probability distributions of $K$ and $L$. 
\blu As may easily be checked, \blk the mean square deviation between  $K$ and $L$ can be calculated 
via the Margenau-Hill distribution as \blu \cite{oz} \blk
\begin{equation} \label{rs}
\langle(\hat{K}-\hat{L})^2\rangle = \sum_{k,l} (\kappa_k-\lambda_l)^2\, p_{\rm MH}(k,l).
\end{equation}
Further,   
the Margenau-Hill distribution itself can be determined   experimentally, via \blu a \blk  weak measurement of  the individual $K_k$ projectors  postselected on a  strong  measurement of  $L$  \cite{lund, Wis03}.  Recalling that $\epsilon(A_{\rm est})=\langle (\hat{A}-\hat{A}_{\rm est})^2\rangle^{1/2}$, this allows the inaccuracies of joint estimates of $A$ and $B$ to be experimentally determined from weak measurements of  the eigenprojectors of $A$ and $B$ (in separate experiments),  postselected on the joint measurement.  This approach also applies if $L$ is replaced by any positive-operator-valued measure $\{\hat{L}_l\}$ having assigned outcome values $\{\lambda_l\}$ \cite{lund}.

Our  generalization 
 of the Lund-Wiseman method  is based on the observation that determining the inaccuracies \blk  requires only that the \blu appropriate \blk Margenau-Hill distributions can be \blu experimentally obtained  in {\it some} manner. \blk   In particular, this can be possible without having to make a weak measurement  \blu --- with \blk the advantage of being simpler to implement experimentally, although at the cost of increasing the inaccuracies.  Further, for our particular experiment, only one \blu semiweak \blk measurement rather than two weak measurements is needed to obtain $\epsilon(X_{\rm est})$ and $\epsilon(Y_{\rm est})$.

In particular, suppose that  in place of a weak measurement of the projector $\hat{K}_k$, the system undergoes a measurement process ${\cal X}$, 
describable by a set of measurement operators $\{\hat{M}_m\}$ \cite{WisMil10},   
such that each $\hat{K}_k$ can be written in the form   $\hat{K}_k=\sum_m \alpha^{(k)}_m \hat{M}_m^\dagger \hat{M}_m$.
\blu Hence, $\langle \hat{K}_k\rangle=\sum_m p(m)\alpha^{(k)}_m$, implying that \blk the coefficients $\alpha^{(k)}_m$ are the contextual values of $K_k$ for the measurement context defined by ${\cal X}$ \cite{dress1}.  It follows that the joint probability distribution for outcome $m$ from measurement ${\cal X}$ and $l$ from a \blu subsequent \blk
measurement of $L$  is  
\begin{equation} \label{pms}
  p(m,l) = \langle \hat{M}_m^\dagger \hat{L}_l\hat{M}_m\rangle, 
\end{equation}
and that the Margenau-Hill distribution for $K$ and $L$ is
\begin{equation} \label{pmh}
  p_{\rm MH}(k,l) = \frac{1}{2} \sum_m \alpha^{(k)}_m \langle \hat{M}_m^\dagger \hat{M}_m\hat{L}_l + \hat{L}_l\hat{M}_m^\dagger \hat{M}_m \rangle. 
\end{equation}
We have found that in many cases 
one can choose ${\cal X}$  such that the Margenau-Hill distribution (\ref{pmh}) can be directly obtained  either  from the measured joint distribution (\ref{pms})  or from (an assumed) theoretical description  of $L$.  In all such cases the corresponding mean square deviation $\langle\,(\hat{K}-\hat{L})^2\,\rangle$ can  be evaluated via Eq.~(\ref{rs}).   

We note that while this method of determining inaccuracies assumes the initial measurement process ${\cal X}$ is describable by a set of measurement operators $\{\hat{M}_m\}$, it can be extended to general ${\cal X}$ via an extension of the definition of contextual values, as will be discussed elsewhere. 

\blu
\subsection{Example: $\epsilon(X_{\rm est})$}

For example, if    $[\hat{L}_l,\hat{M}_m]=0$  then Eqs.~(\ref{pms}) and (\ref{pmh}) yield the \blu simple \blk relation $p_{\rm MH}(k,l) = \sum_m \alpha^{(k)}_m p(m,l)$.  \blu This can be applied in our experiment, noting that the  measurement process ${\cal X}$ in Fig.~1 only acts on the first qubit, so that $[\hat{M}_m,\hat{W}_w]$ = 0. Since the contextual values of the projectors $\hat{X}_\pm=(1\pm\hat{X})/2$ follow from the contextual values of ${\cal X}$ in the previous section as $\alpha^{(\pm)}_m=\frac{1}{2}(1\pm \xi_m)$,  the inaccuracy of any estimate of the $X$ polarisation of the form $X_{\rm est}=f(W)$ follows via Eq.~(\ref{rs}) as 
\begin{eqnarray*} 
\epsilon(X_{\rm est})^2 &=&\langle [\hat{X}\otimes \hat{\mathbbm{1}} - \hat{\mathbbm{1}}\otimes f(\hat{W})]^2\rangle\\
&=& \sum_{x,w} [x-f(w)]^2 p_{\grn \rm MH}(x,w)\\
&=& \sum_{m,x,w} [x-f(w)]^2 \alpha^{(x)}_m p(m,w) ,
\end{eqnarray*}
with $x=\pm 1$.  Substituting for $\alpha^{(x)}_m$, and noting that $p(m,w)=\sum_y p(m,y,w)$ by definition, yields the expression in Eq.~(6) of the main text.

\subsection{Example: $\epsilon(Y_{\rm est})$}

\blk
As a second example,  \blu note that applying ${\cal X }$ before the measurement of $Y$, as in Fig.~1(b) of the main text, is equivalent to measuring the positive operator valued measure ${\cal Y}=\{\hat{\Upsilon}_\pm\}$ on the initial state, with $\hat{\Upsilon}_\pm= \sum_m \hat{M}_m^\dagger \hat{Y}_\pm \hat{M}_m$, where \red $\hat{Y}_\pm:=(1\pm\hat{Y})/2$ \blk denotes the projectors corresponding to $\hat{Y}$ \cite{WisMil10}.   Using the forms of $\hat{M}_r$ and $\hat{M}_t$ in the previous section, this simplifies to $\hat{\Upsilon}_\pm=\frac{1}{2} \pm\frac{1}{2}(1-\kappa) \hat{Y}$, as noted in the main text (corresponding to white noise of magnitude $\kappa$), where  $\kappa$ is a function of $r_H$ and $r_V$ related to the strength of the intermediate measurement process ${\cal X}$.  
The Margenau-Hill distribution for $Y$ and ${\cal Y}$ follows directly from Eq.~(\ref{mh}) as 
\[ p_{\grn \rm MH}(y,y')=\frac{1}{2}\langle \hat{Y}_{y}\hat{\Upsilon}_{y'} + \hat{\Upsilon}_{y'}\hat{Y}_{y}\rangle=[\kappa/2+(1-\kappa)\delta_{yy'}]\langle \hat{Y}_y\rangle, \] 
which combined with Eq.~(\ref{rs}) yields a corresponding inaccuracy 
$\epsilon(Y_{\rm est})^2 = \sum_{y,y'=\pm 1} [y-y']^2 \, p_{\grn \rm MH}(y,y') =2\kappa$,
as given \gold in 
\blk the main text.

\blk


\begin{thebibliography}{99}
\bibitem{bohr1} N. Bohr, {\it  Atomic Physics and Human Knowledge} (Wiley, New
York, 1958), pp. 32-66.
\bibitem{epr}  A. Einstein, B. Podolsky, and N. Rosen, {Phys. Rev.} {\bf 47}, 777
(1935).
\bibitem{bohrepr} N. Bohr, {Phys. Rev.} {\bf 48}, 696 (1935).
\bibitem{ozawadist} M. Ozawa,  {Phys. Rev. A} {\bf 67}, 042105 (2003).
\bibitem{ozawaannphys} M. Ozawa,  {Ann. Phys. (Amsterdam)} {\bf 311}, 350 (2004).
\bibitem{ozawanat}   J. Erhart, S. Sponar, G. Sulyok, G. Badurek, M. Ozawa,
and Y. Hasegawa,   {Nature Phys.} {\bf 8} 185 (2012).
\bibitem{rozema}   L. A. Rozema, A. Darabi, D. H. Mahler, A. Hayat, Y. Soudagar, and A. M. Steinberg,  {Phys. Rev. Lett.} {\bf 109}, 100404 (2012). 
\bibitem{vantrees}  H. L. van Trees,
{\it  Detection, Estimation and Modulation Theory Part 1}
{(Wiley, New York, 2001)}, Sec.~{2.4}.
\bibitem{ozawa1} M. Ozawa, in {\it  Quantum Aspects of Optical Communications},
(eds C. Bendjaballah, O. Hirota, and S. Reynaud) {\it  Lecture Notes in Physics} {\bf 378}  (Springer, Berlin, 1991), pp. 3-17.
\blu
\bibitem{note}  
Note the estimates referred to here are of observables, and hence are represented by operators.  They should not be confused with estimates of classical parameters of quantum states, the theory of which was pioneered by Helstrom and Holevo \cite{helhol}.
\bibitem{ak} E. Arthurs and J. L. Kelly, Jr., {Bell Syst. Tech. J.} {\bf 44}, 725
(1965).
\bibitem{ag}  E. Arthurs and M. S. Goodman, {Phys. Rev. Lett.} {\bf 60}, 2447
(1988).
\bibitem{ishi} S. Ishikawa, Rep. Math. Phys. {\bf 29}, 257 (1991).
\bibitem{woot}  W. K. Wootters and W. H. Zurek, W. H., {Phys. Rev. D} {\bf 19}, 473
(1979). 
\bibitem{mart} H. Martens and W. de Muynck,  {Found. Phys.} {\bf 20}, 357 (1990).
\bibitem{jaeger} G. Jaeger, A. Shimony, and L. Vaidman,  {Phys. Rev. A} {\bf 51}, 54 (1995).
\bibitem{apple}  D. M. Appleby,  {Int. J. Theor. Phys.} {\bf 37}, 1491 (1998).
\bibitem{trif}  A. Trifonov, G. Bj\"ork, and J. S\"oderholm,  {Phys. Rev. Lett.} {\bf 86}, 4423 (2001).
\bibitem{busch} P. Busch and C. R. Shilladay,  {Phys. Rev. A} {\bf 68}, 034102 (2003).
\bibitem{watanabe} Y. Watanabe, T. Sagawa and M. Ueda, {Phys. Rev. A} {\bf 84}, 042121 (2011).
\bibitem{hall}  M. J. W. Hall, {Phys. Rev. A} {\bf 69}, 052113 (2004).
\bibitem{eur} M. J. W. Hall, {Phys. Rev. A} {\bf 64}, 052103 (2001).
\bibitem{ozawa} M. Ozawa, Int. J. Quantum Inf. {\bf 01}, 569 (2003); \blk M. Ozawa,  {Phys. Lett. A} {\bf 320}, 367 (2004).
\bibitem{lund}  A. P. Lund and H. M. Wiseman,  {New. J Phys.} {\bf 12}, 093011 (2010).
\bibitem{dress1} J. Dressel, S. Agarwal, and A. N. Jordan,  {Phys. Rev. Lett.} {\bf 104},  240401 (2010).
\bibitem{supp} See appended Supplemental Material for derivation of theoretical results and further experimental details.
\bibitem{WisMil10} \blk   {H. M. Wiseman  and G. J. Milburn}, {\it  Quantum Measurement and Control}
 {(Cambridge University Press, Cambridge, 2010)}.  
\bibitem{tom} A. G. White, A. Gilchrist, G. J. Pryde, J. L. O'Brien, M. J. Bremner, and N. K. Langford,   {JOSA B} {\bf 24}, 172 (2007).
\bibitem{pryde05} G. J. Pryde, J. L.  O'Brien, A. G. White, T. C. Ralph, and H. M. Wiseman,  {Phys. Rev. Lett.} {\bf 94}, 220405 (2005).
\bibitem{dress2} J. Dressel, C. J. Broadbent, J. C. Howell, and A. N. Jordan, {Phys. Rev. Lett.} {\bf 106}, 040402 (2011).
\bibitem{buscemi} F. Buscemi,   {Phys. Rev. Lett.} {\bf 108}, 200401 (2012).
\bibitem{dense} C. H. Bennett and S. J. Wiesner, {Phys. Rev. Lett.} {\bf 69}, 2881 (1992).
\bibitem{crypt} N. Gisin, G. Ribordy, W. Tittel, and H. Zbinden,  {Rev. Mod. Phys.} {\bf 74}, 145 (2002). \blk
\bibitem{helhol}  {C. W. Helstrom}, {\it Quantum Detection and Estimation Theory}
 {(Academic Press, New York, 1976)};  A. S.  Holevo, {\it Probabilistic and Statistical Aspects of Quantum Theory} (North-Holland, Amsterdam, 1982). 






\bibitem{branc} C. Branciard, Proc. Nat. Acad. Sci., {\bf 110} 6742 (2013).
\blk
\bibitem{source1} R. Rangarajan, M. Goggin, and P. Kwiat, {Opt. Express} {\bf 17}, 18920 (2009).
\bibitem{source2}  P. G. Kwiat, E. Waks, A. G. White, I. Appelbaum, and P. H. Eberhard,  {Phys. Rev. A} {\bf 60}, R773 (1999).

\bibitem{mh} H. Margenau and R. N. Hill, {Prog. Theor. Phys.} {\bf 26}, 722 (1961).
\blu
\bibitem{oz} M. Ozawa, Phys. Lett. A {\bf 335}, 11 (2005).
\blk
\bibitem{Wis03} H. M. Wiseman, 	 {Phys. Lett. A} {\bf 311}, 285 (2003). \blk   
\bibitem{kraus} K. Kraus, {\it  States, Effects and Operations} (Springer, Berlin, 1983).

\end{thebibliography}
\end{document}